\documentstyle[epsfig]{elsart}
\def\bk{{\bf k}}

\begin{document}
\begin{frontmatter}
\title{Coulomb corrections to the three-body correlation function in high-energy 
heavy ion reactions}
\author[Mainz]{E. O.\ Alt\thanksref{erwin},} 
\author[KFKI]{T. Cs{\"o}rg\H o\thanksref{tamas},}
\author[Lund]{B. L{\"o}rstad\thanksref{bengt},}
\author[Lund]{J. Schmidt-S{\o}rensen\thanksref{janus}}
\address[Mainz]{Institut f\"ur Physik, Universit\"at Mainz, D-55099 Mainz, Germany}
\address[KFKI]{MTA KFKI RMKI, H-1525 Budapest 114, POB 49, Hungary}
\address[Lund]{Physics Department, Lund University, S-221 00 Lund, POB 118, Sweden}

\thanks[erwin]{Email: alt@dipmza.physik.uni-mainz.de}
\thanks[tamas]{Email: csorgo@sunserv.kfki.hu}
\thanks[bengt]{Email: bengt@quark.lu.se}
\thanks[janus]{Email: janus@quark.lu.se}

\begin{abstract}
Starting from an asymptotically correct three-body Coulomb wave-function,
we determine the effect of Coulomb final state interaction on the 
three-particle Bose-Einstein correlation function of similarly charged
particles. We numerically estimate that the 
Riverside approximation is not precise enough to 
determine the three-body Coulomb correction factor
in the correlation function, 
if the characteristic HBT radius parameter is 5 - 10 fm,
which is the range of interest in high-energy heavy ion physics.
\end{abstract}
\end{frontmatter}

\section{Introduction}

One of the most important tasks of high energy heavy-ion studies is to prove the existence
of the elusive quark-gluon plasma and to study the properties of this predicted new state 
of matter\cite{muller}.
Hanbury-Brown Twiss (HBT) 
interferometry\cite{hanbury,Boa90a,Lorstad,Baym,Weiner} 
of identical particles has become an important tool
as it can be used to measure the evolving geometry of the interaction region.
The quantitative interpretation of the HBT-results depends however critically on the understanding
of the r{\^o}le of the Coulomb interaction between the detected particles as well as
on the influence exerted on the selected system by the remaining particles\cite{BaymM,humanic,barz}.
In the attempts to create the quark-gluon plasma we  use higher
and higher energies and larger and larger colliding nuclei. In these collisions
the multiplicity of particles gets very large and the importance of the HBT-study
of more than two particles will grow. Recently, much work \cite{heinz,axel,bo} 
has been devoted to what can be
learned from the  system of three charged pions, however without paying sufficient attention on how the
Coulomb interaction might change the predictions. 
In the few existing experimental studies
on the HBT effect in the three-charged pion system, see \cite{bengt} and references given
 there, Coulomb effects have been taken into account in
 the so-called
Riverside approximation \cite{riverside},  or sometimes in more elaborated versions of it\cite{Cramer}.
However, these schematic conceptual ansaetze are acceptable only because a more accurate
multiparticle Coulomb interaction treatment has not been available.

In this Letter we propose a proper treatment of the Coulomb interaction
between three charged particles. It is based on non-relativistic three-particle scattering 
theory, assuming that their velocities relative to each other are non-relativistic which, 
however, is just the kinematic situation where the HBT study of three particles is of interest. 
We, moreover, show that the Riverside  approximation follows as a special case when the source 
of particle creation
becomes very small, in the same way as the Gamow 
penetration factor is the limit
for the Coulomb correction in the two-particle correlation function. An estimate
 of the magnitude of the corrections for source sizes to be expected at RHIC and LHC is presented using
current data from NA44 experiment on S-Pb collisions at CERN SPS.

\section{Coulomb corrections to three-body correlations}

\subsection{Three-charged particle wave function} 
In order to treat correctly the Coulomb corrections to the three-body correlation function 
knowledge of the three-particle Coulomb wave function is required. As mentioned in the 
Introduction, we restrict ourselves to the case that the transverse momenta of the three 
particles in the final state are small enough to make a nonrelativistic approach applicable.
 Hence the problem consists in finding the solution of the three-charged particle 
 Schr\"odinger equation when all three particles are in the continuum. 

Consider three distinguishable particles with masses 
$m_{i}$ and charges $e_{i},\,i = 1,2,3$. Let ${\rm {\bf {x}}}_{i}$ and ${\rm {\bf {k}}}_{i}$
 denote the coordinate and momentum (three-)vectors, respectively, of particle $i$. 
 From these we construct in the usual manner the relative coordinate ${\rm {\bf {r}}}_{ij} =
  {\rm {\bf {x}}}_{i} - {\rm {\bf {x}}}_{j}$ and the relative momentum ${\rm {\bf {k}}}_{ij}
   = (m_j {\rm {\bf {k}}}_{i} - m_i {\rm {\bf {k}}}_{j})/{(m_{i} +m_{j})}$ between particles
    $i$ and $j$, the corresponding reduced mass being $\mu_{ij} = m_{i}m_{j}/(m_{i} +m_{j})$.

The three-particle Schr\"odinger equation reads
\begin{eqnarray}
\left\{ H_0 + \sum_{i < j=1}^{3} V_{ij} - E \right\} \Psi^{(+)}_{ {\rm {\bf {k}}}_{1} {\rm 
{\bf {k}}}_{2} {\rm {\bf {k}}}_{3} }({\rm {\bf {x}}}_{1},{\rm {\bf {x}}}_{2},{\rm 
{\bf {x}}}_{3}) = 0, \label{3se}
\end{eqnarray}
where
\begin{eqnarray}
E = \sum_{i=1}^{3}\frac{k_{i}^2}{2m_{i}} > 0
\end{eqnarray}
is the total kinetic energy for three particles in the continuum. $H_0$ is the three-free 
particle Hamilton operator and
\begin{eqnarray}
V_{ij}({\rm {\bf {r}}}_{ij}) = V_{ij}^S({\rm {\bf {r}}}_{ij}) + V_{ij}^C({\rm {\bf {r}}}_{ij}) 
\end{eqnarray}
the interaction potential between particles $i$ and $j$, consisting of a sum of a strong 
short-range ($V_{ij}^S$) plus the long-range Coulomb interaction ($V_{ij}^C({\rm {\bf {r}}}_{ij})
 = e_i e_j/r_{ij}$). 

The exact numerical solution of the Schr\"odinger equation (\ref{3se}) 
for $E>0$ is beyond present means, partly for principal and partly 
for practical reasons. For a brief discussion of the related difficulties 
see \cite{a98b}. But, at least in all asymptotic regions of the three-particle 
configuration space is the form of its solution nowadays 
known analytically \cite{r72,am92}. It is simplest in the asymptotic 
region usually denoted by $\Omega _{0}$ and characterized by the 
fact that - roughly speaking - all three interparticle distances 
become uniformly large, i.e., $r_{12}, r_{23}, r_{31} \to \infty$ (for 
a precise definition of the various asymptotic regions see \cite{am92}). 

A convenient and widely used representation of a three-charged particle wave function which 
coincides in $\Omega _{0}$ with the correct asymptotic wave function is given by \cite{m77,bbk89}
\begin{eqnarray} 
\Psi^{(+)}_{ {\rm {\bf {k}}}_{1} {\rm {\bf {k}}}_{2} {\rm {\bf {k}}}_{3} }({\rm 
{\bf {x}}}_{1},{\rm {\bf {x}}}_{2},{\rm {\bf {x}}}_{3}) \; \stackrel{\Omega _{0}}
{\sim} \; {\psi}_{ {\rm {\bf {k}}}_{12}}^{C(+)}({\rm {\bf {r}}}_{12}) {\psi}_{ 
{\rm {\bf {k}}}_{23}}^{C(+)}({\rm {\bf {r}}}_{23}) {\psi}_{ {\rm {\bf {k}}}_{31}}^{C(+)}
({\rm {\bf {r}}}_{31}). \label{aswfn}
\end{eqnarray} 
Here, ${\psi}_{ {\rm {\bf {k}}}_{ij}}^{C(+)}({\rm {\bf {r}}}_{ij})$ is 
the continuum solution of the two-body Coulomb Schr\"odinger equation 
\medskip
\begin{eqnarray} 
\left\{ - \frac{\Delta_{{\rm {\bf{r}}}_{ij}}}{2 \mu_{ij}} + V_{ij}^C({\rm {\bf{r}}}_{ij}) - 
\frac {k_{ij}^{2}}{2 \mu_{ij}} \right\} {\psi}_{ {\rm{\bf{k}}}_{ij}}^{C(+)}({\rm {\bf{r}}}_{ij})
 = 0, \label{2cse} 
\end{eqnarray}
describing the relative motion of the two particles $i$ and $j$ with 
energy $k_{ij}^{2}/2 \mu_{ij}$. The explicit solution of (\ref{2cse}) is known,
\medskip
\begin{eqnarray} 
{\psi}_{ {\rm {\bf {k}}}_{ij}}^{C(+)}({\rm {\bf {r}}}_{ij}) &=& N _{ij} e^{ i {\rm 
{\bf {k}}}_{ij} \cdot {\rm {\bf {r}}}_{ij}} F[- i \eta _{ij}, 1; i ( { k}_{ij} r_{ij}- 
{\rm {\bf {k}}}_{ij} \cdot {\rm {\bf {r}}}_{ij})], \label{2pwfn}
\end{eqnarray} 
with $N_{ij} = e^{- \pi \eta_{ij}/2}\, 
\Gamma (1 + i \eta_{ij}), $ and 
$\eta_{ij}= \frac{e_{i} e_{j}\mu_{ij}}{ k_{ij}}$ being the appropriate 
Coulomb (Sommerfeld) parameter.
 $F[a,b;x]$ is the confluent hypergeometric function and 
$\Gamma (x)$ the Gamma function. When writing the asymptotic 
solution of (\ref{3se}) in the form (\ref{aswfn}), use has been made of 
the fact that in $\Omega _{0}$, 
the short-range part $V_{ij}^S$ of the two-particle interactions 
can be neglected and, hence, only the Coulomb potentials survive.

Let us add a few comments. \\
(i) It is to be emphasized that in using a wave function of the
type (\ref{aswfn}), the three-body system is considered as a sum of 
three
noninteracting two-body systems (each on its two-body energy shell).
That is, neither correlations between the motions of the three particle
pairs nor off-the-two-body-shell effects are included, which obviously
can be true at most for asymptotic particle separations
(in fact, it holds true in, and only in, $\Omega_{0}$). \\
(ii) The wave function (\ref{aswfn}) provides 
a well-defined prescription of how to go to arbitrary, in particular small, values 
of the relative coordinates. Though, it ceases to be solution 
of the Schr\"odinger equation (\ref{3se}) for non-asymptotic 
values of the relative coordinates; nevertheless, it 
has proved to be rather successful in describing $(e,2e)$ 
processes in atomic physics. But it should be kept in mind that 
the extrapolation out of the region $\Omega_{0}$ 
implied by (\ref{aswfn}) is highly non-unique. In fact, several other 
ansaetze which, of course, coincide asymptotically in $\Omega_{0}$ 
with (\ref{aswfn}) have been, and are still being, developed. \\
(iii) We mention that ansaetze for three-charged particle wave 
functions have been proposed which are correct in all asymptotic 
regions (see, e.g., \cite{am92}). As is to be expected, 
they contain correlations between the motions of the three particles. 
However, their form is rather more complicated and, thus, will not be 
used for the present investigation. For some more details and for 
further references see \cite{a98a}.

Hence, in the present investigation we assume the three-charged particle 
wave function to be given in all of three-body configuration space as
\begin{eqnarray}
\tilde \Psi^{(+)}_{ {\rm {\bf {k}}}_{1} {\rm {\bf {k}}}_{2} {\rm {\bf {k}}}_{3} }({\rm 
{\bf {x}}}_{1},{\rm {\bf {x}}}_{2},{\rm {\bf {x}}}_{3}) \;:= \; {\cal N} {\psi}_{ {\rm 
{\bf {k}}}_{12}}^{C(+)}({\rm {\bf {r}}}_{12}) {\psi}_{ {\rm {\bf {k}}}_{23}}^{C(+)}({\rm 
{\bf {r}}}_{23}) {\psi}_{ {\rm {\bf {k}}}_{31}}^{C(+)}({\rm {\bf {r}}}_{31}). \label{aswfn1}
\end{eqnarray}
Here, ${\cal N}$ is an (undetermined) overall normalization constant.

\subsection{Specialization to three identical bosons}

Given a general three-particle wave function $\Psi^{(+)}_{ {\rm {\bf {k}}}_{1} {\rm 
{\bf {k}}}_{2} {\rm {\bf {k}}}_{3} }({\rm {\bf {x}}}_{1},{\rm {\bf {x}}}_{2},{\rm 
{\bf {x}}}_{3}) $ it is a simple task to specialize to various interesting situations. 
As a particular application we treat explicitly only the case of three identical bosons 
with unit charges of magnitude $|e|$ and mass $m$. The properly symmetrized wave function 
is
\begin{eqnarray}
\Psi^{(+){\cal S}}_{ {\rm {\bf {k}}}_{1} {\rm {\bf {k}}}_{2} {\rm {\bf {k}}}_{3} }({\rm 
{\bf {x}}}_{1},{\rm {\bf {x}}}_{2},{\rm {\bf {x}}}_{3}) &=& \frac{1}{\sqrt{6}} 
\left\{ \Psi^{(+)}_{ {\rm {\bf {k}}}_{1} {\rm {\bf {k}}}_{2} {\rm {\bf {k}}}_{3} }
({\rm {\bf {x}}}_{1},{\rm {\bf {x}}}_{2},{\rm {\bf {x}}}_{3}) + \Psi^{(+)}_{ {\rm 
{\bf {k}}}_{1} {\rm {\bf {k}}}_{2} {\rm {\bf {k}}}_{3} }({\rm {\bf {x}}}_{1},{\rm 
{\bf {x}}}_{3},{\rm {\bf {x}}}_{2}) \right. \nonumber \\
&& \left.+ \Psi^{(+)}_{ {\rm {\bf {k}}}_{1} {\rm {\bf {k}}}_{2} 
{\rm {\bf {k}}}_{3} }({\rm {\bf {x}}}_{2},{\rm {\bf {x}}}_{1},{\rm {\bf {x}}}_{3}) + \Psi^{(+)}_{ {\rm {\bf {k}}}_{1} {\rm {\bf {k}}}_{2} {\rm {\bf {k}}}_{3} }
({\rm {\bf {x}}}_{2},{\rm {\bf {x}}}_{3},{\rm {\bf {x}}}_{1})\right. \nonumber \\
&& \left.+ \Psi^{(+)}_{ {\rm 
{\bf {k}}}_{1} {\rm {\bf {k}}}_{2} {\rm {\bf {k}}}_{3} }({\rm {\bf {x}}}_{3},{\rm 
{\bf {x}}}_{1},{\rm {\bf {x}}}_{2})+ \Psi^{(+)}_{ {\rm {\bf {k}}}_{1} {\rm {\bf {k}}}_{2} 
{\rm {\bf {k}}}_{3} }({\rm {\bf {x}}}_{3},{\rm {\bf {x}}}_{2},{\rm {\bf {x}}}_{1})\right\}.
\end{eqnarray}

By choosing for $\Psi^{(+)}_{ {\rm {\bf {k}}}_{1} {\rm {\bf {k}}}_{2} {\rm {\bf {k}}}_{3} }
({\rm {\bf {x}}}_{1},{\rm {\bf {x}}}_{2},{\rm {\bf {x}}}_{3}) $ the - at least asymptotically 
in $\Omega _{0}$ correct - ansatz 
\begin{eqnarray}
\Psi^{(+)}_{ {\rm {\bf {k}}}_{1} {\rm {\bf {k}}}_{2} {\rm {\bf {k}}}_{3} }({\rm 
{\bf {x}}}_{1},{\rm {\bf {x}}}_{2},{\rm {\bf {x}}}_{3}) = \tilde \Psi^{(+)}_{ {\rm 
{\bf {k}}}_{1} {\rm {\bf {k}}}_{2} {\rm {\bf {k}}}_{3} }({\rm {\bf {x}}}_{1},{\rm 
{\bf {x}}}_{2},{\rm {\bf {x}}}_{3}),
\end{eqnarray}
we find for the approximate symmetrized wave function the form
\begin{eqnarray}
\tilde\Psi^{(+){\cal S}}_{ {\rm {\bf {k}}}_{1} {\rm {\bf {k}}}_{2} 
{\rm {\bf {k}}}_{3} }({\rm {\bf {x}}}_{1},{\rm {\bf {x}}}_{2},
{\rm {\bf {x}}}_{3}) &=& 
\frac{{\cal N}}{\sqrt{6}} 
\left\{ 
{\psi}_{ {\rm {\bf {k}}}_{12}}^{C(+)}({\rm {\bf {r}}}_{12}) 
{\psi}_{ {\rm {\bf {k}}}_{23}}^{C(+)}({\rm {\bf {r}}}_{23}) 
{\psi}_{ {\rm {\bf {k}}}_{31}}^{C(+)}({\rm {\bf {r}}}_{31}) 
 \right. \nonumber \\
&+& 
\left. 
 {\psi}_{ {\rm {\bf {k}}}_{12}}^{C(+)}({\rm {\bf {r}}}_{13}) 
 {\psi}_{ {\rm {\bf {k}}}_{23}}^{C(+)}({\rm {\bf {r}}}_{32}) 
 {\psi}_{ {\rm {\bf {k}}}_{31}}^{C(+)}({\rm {\bf {r}}}_{21})
 \right. \nonumber \\
&+& 
\left. 
{\psi}_{ {\rm {\bf {k}}}_{12}}^{C(+)}({\rm {\bf {r}}}_{21}) 
{\psi}_{ {\rm {\bf {k}}}_{23}}^{C(+)}({\rm {\bf {r}}}_{13}) 
{\psi}_{ {\rm {\bf {k}}}_{31}}^{C(+)}({\rm {\bf {r}}}_{32}) 
 \right. \nonumber \\
&+& 
\left. 
 {\psi}_{ {\rm {\bf {k}}}_{12}}^{C(+)}({\rm {\bf {r}}}_{23}) 
 {\psi}_{ {\rm {\bf {k}}}_{23}}^{C(+)}({\rm {\bf {r}}}_{31}) 
 {\psi}_{ {\rm {\bf {k}}}_{31}}^{C(+)}({\rm {\bf {r}}}_{12})
\right. \nonumber \\
&+& 
\left. 
{\psi}_{ {\rm {\bf {k}}}_{12}}^{C(+)}({\rm {\bf {r}}}_{31}) 
{\psi}_{ {\rm {\bf {k}}}_{23}}^{C(+)}({\rm {\bf {r}}}_{12}) 
{\psi}_{ {\rm {\bf {k}}}_{31}}^{C(+)}({\rm {\bf {r}}}_{23}) 
 \right. \nonumber \\
&+& 
\left. 
 {\psi}_{ {\rm {\bf {k}}}_{12}}^{C(+)}({\rm {\bf {r}}}_{32}) 
 {\psi}_{ {\rm {\bf {k}}}_{23}}^{C(+)}({\rm {\bf {r}}}_{21}) 
 {\psi}_{ {\rm {\bf {k}}}_{31}}^{C(+)}({\rm {\bf {r}}}_{13}) 
 \right\}. \nonumber \\
\label{wfnsym}
\end{eqnarray}

The case that all three particles are uncharged follows from (\ref{wfnsym}) by substituting 
for the Coulomb wave functions ${\psi}_{ {\rm {\bf {k}}}_{ij}}^{C(+)}({\rm {\bf {r}}}_{ij})$ 
the plane waves $e^{ i {\rm {\bf {k}}}_{ij} \cdot {\rm {\bf {r}}}_{ij}} $. 
We denote the corresponding symmetrized three-uncharged particle wave function by 
$\tilde\Psi^{(0){\cal S}}_{ {\rm {\bf {k}}}_{1} {\rm {\bf {k}}}_{2} {\rm {\bf {k}}}_{3} }
({\rm {\bf {x}}}_{1},{\rm {\bf {x}}}_{2},{\rm {\bf {x}}}_{3})$:
\begin{eqnarray}
\tilde\Psi^{(0){\cal S}}_{ {\rm {\bf {k}}}_{1} {\rm {\bf {k}}}_{2} {\rm {\bf {k}}}_{3} }({\rm 
{\bf {x}}}_{1},{\rm {\bf {x}}}_{2},{\rm {\bf {x}}}_{3})&:&= \frac{{\cal N}_0}{\sqrt{6}} 
\left\{ e^{ i {\rm {\bf {k}}}_{12} \cdot {\rm {\bf {r}}}_{12}} e^{ i {\rm {\bf {k}}}_{23} 
\cdot {\rm {\bf {r}}}_{23}} e^{ i {\rm {\bf {k}}}_{31} \cdot {\rm {\bf {r}}}_{31}} +
 e^{ i {\rm {\bf {k}}}_{12} \cdot {\rm {\bf {r}}}_{13}} e^{ i {\rm {\bf {k}}}_{23} \cdot 
 {\rm {\bf {r}}}_{32}} e^{ i {\rm {\bf {k}}}_{31} \cdot {\rm {\bf {r}}}_{21}} \right. \nonumber \\
&& \left.+  e^{ i {\rm {\bf {k}}}_{12} \cdot {\rm {\bf {r}}}_{21}} e^{ i {\rm {\bf {k}}}_{23} 
\cdot {\rm {\bf {r}}}_{13}} e^{ i {\rm {\bf {k}}}_{31} \cdot {\rm {\bf {r}}}_{32}} + 
e^{ i {\rm {\bf {k}}}_{12} \cdot {\rm {\bf {r}}}_{23}} e^{ i {\rm {\bf {k}}}_{23} \cdot 
{\rm {\bf {r}}}_{31}} e^{ i {\rm {\bf {k}}}_{31} \cdot {\rm {\bf {r}}}_{12}} \right. \nonumber \\
&& \left.+  e^{ i {\rm {\bf {k}}}_{12} \cdot {\rm {\bf {r}}}_{31}} e^{ i {\rm {\bf {k}}}_{23} 
\cdot {\rm {\bf {r}}}_{12}} e^{ i {\rm {\bf {k}}}_{31} \cdot {\rm {\bf {r}}}_{23}} +
 e^{ i {\rm {\bf {k}}}_{12} \cdot {\rm {\bf {r}}}_{32}} e^{ i {\rm {\bf {k}}}_{23} 
 \cdot {\rm {\bf {r}}}_{21}} e^{ i {\rm {\bf {k}}}_{31} \cdot {\rm {\bf {r}}}_{13}} 
 \right\}, \label{wfn0sym}
\end{eqnarray}
where ${\cal N}_0$ is the appropriate normalisation constant. 

A simple approximation to (\ref{wfnsym}) which incorporates at least part of 
the Coulomb effects is obtained by substituting for each of the two-particle 
Coulomb wave functions 
${\psi}_{ {\rm {\bf {k}}}_{ij}}^{C(+)}({\rm {\bf {r}}}_{ij})$ the term 
$e^{ i {\rm {\bf {k}}}_{ij} \cdot {\rm {\bf {r}}}_{ij}} N_{ij}$, 
i.e., by neglecting the hypergeometric function part in the exact solution 
(\ref{2pwfn}) (this is justified if, for all pair of indices $(ij)=(12),(23),(31)$, 
the arguments $({k}_{ij} r_{ij}- {\rm {\bf {k}}}_{ij} \cdot {\rm {\bf {r}}}_{ij})$ of 
the hypergeometric functions are sufficiently small, as it happens, e.g.,  for small relative 
distances). This leads to
\begin{eqnarray}
\tilde\Psi^{(+){\cal S}}_{ {\rm {\bf {k}}}_{1} {\rm {\bf {k}}}_{2} {\rm {\bf {k}}}_{3} }
({\rm {\bf {x}}}_{1},{\rm {\bf {x}}}_{2},{\rm {\bf {x}}}_{3}) \approx N_{12}N_{23}N_{31} 
\Psi^{(0){\cal S}}_{ {\rm {\bf {k}}}_{1} {\rm {\bf {k}}}_{2} {\rm {\bf {k}}}_{3} }({\rm 
{\bf {x}}}_{1},{\rm {\bf {x}}}_{2},{\rm {\bf {x}}}_{3}).
\end{eqnarray}
For the probability density we arrive at 
\begin{eqnarray}
\left|\tilde\Psi^{(+){\cal S}}_{ {\rm {\bf {k}}}_{1} {\rm {\bf {k}}}_{2} {\rm 
{\bf {k}}}_{3} }({\rm {\bf {x}}}_{1},{\rm {\bf {x}}}_{2},{\rm {\bf {x}}}_{3})\right|^2 
\approx G_{12}G_{23}G_{31} \left|\tilde\Psi^{(0){\cal S}}_{ {\rm {\bf {k}}}_{1} {\rm 
{\bf {k}}}_{2} {\rm {\bf {k}}}_{3} }({\rm {\bf {x}}}_{1},{\rm {\bf {x}}}_{2},{\rm 
{\bf {x}}}_{3})\right|^2,
\end{eqnarray}
where the Gamov factors $G_{ij} \equiv \left| N_{ij} \right|^2$ have 
been introduced. This is nothing but the well known 
Riverside approximation which consequently finds some theoretical 
justification within the framework of the exact nonrelativistic 
three-body scattering theory as a kind of lowest-order approximation.
But this justification is rather weak. For, in order to "derive" the Riverside approximation, 
the wave function (\ref{aswfn1}) or (\ref{wfnsym}) is used in the region 
$r_{12}, r_{23}, r_{31} \approx 0$ where 
it is certainly a much poorer approximation to the exact solution of (\ref{3se}) than at the larger 
values of the relative particle separations as applied in the following.
Thus, a better justified expression is obtained by 
using the wave function (\ref{wfnsym}). Nevertheless, 
also in that case an overall factor $N_{12}N_{23}N_{31}$ 
can be extracted from the r.h.s.\ of (\ref{wfnsym}), 
which for the probability density yields again a factor 
$G_{12}G_{23}G_{31}$. Note that
in the context of three-body scattering theory we do not find any support for the 
Coulomb correction factors proposed in \cite{Cramer}.

\subsection{Application to high-energy heavy-ion collisions}

The correlation function measuring the enhanced probability for emission of three identical 
Bose-particles is given by:
\begin{eqnarray}
 C_3({\bf k}_1,{\bf k}_2,{\bf k}_3) & =  &
\frac{\displaystyle N_3({\bf k}_1,{\bf k}_2,{\bf k}_3)}
{\displaystyle N_1({\bf k}_1)\, N_1({\bf k}_2)\, N_1({\bf k}_3)}
\label{corr-func}
\end{eqnarray}
where ${\bf k}_i$ is the three-momentum of particle $i$.  
The three-particle momentum distribution is $ N_3({\bf k}_1,{\bf k}_2,{\bf k}_3)$
and the single-particle momentum distribution 
is denoted by $ N_i({\bf k}_i)$.
 This correlation function
is presently, due to meagre statistics, 
only measured as a function of the Lorentz 
invariant $Q_3$, defined by the relation
\begin{equation}
Q_3^2=k_{12}^2+k_{23}^2+k_{31}^2
\end{equation}
where $k_{ij}=k_i-k_j$. In these equations is $k_i$ the four-momentum vector
defined by the three-momentum ${\bf k}_i$ and the mass of particle $i$, $m_i$.

We can now calculate the Coulomb effect on 
this three-particle correlation function using 

\begin{equation}
K_{Coulomb}(Q_{3})= \frac{\int d^{3}{\bf x}_{1} \rho ({\bf x}_{1}) d^{3}{\bf x}_{2} 
\rho ({\bf x}_{2})
d^{3}{\bf x}_{3} \rho ({\bf x}_{3})
\left|\tilde\Psi^{(+)S}_{ {\rm {\bf {k}}}_{1} {\rm {\bf {k}}}_{2} 
{\rm {\bf {k}}}_{3} }({\rm {\bf {x}}}_{1},{\rm {\bf {x}}}_{2},{\rm {\bf {x}}}_{3})\right|^2}
{\int d^{3}{\bf x}_{1} \rho ({\bf x}_{1}) d^{3}{\bf x}_{2} \rho ({\bf x}_{2})
d^{3}{\bf x}_{3} \rho ({\bf x}_{3})
\left|\tilde\Psi^{(0)S}_{ {\rm {\bf {k}}}_{1} {\rm {\bf {k}}}_{2} {\rm {\bf {k}}}_{3} }
({\rm {\bf {x}}}_{1},{\rm {\bf {x}}}_{2},{\rm {\bf {x}}}_{3})\right|^2}
\label{Kcoul3},
\end{equation}
where $\rho({\bf x_i})$ is the density distribution of the source for
 particle $i$, taken as a Gaussian distribution of width R in all three spatial
  directions, $\rho({\bf x})= \frac{1}{(2\pi R)^{3}} \exp[-(\frac{{\bf x}^2}{2R^2})]$. This 
  formulation makes it possible to estimate the Coulomb effect as a function of the
radius parameter $R$ on the three-particle correlation function,
 and to compare this estimate with that calculated by 
   means of the Riverside approximation through 
   $K_{Coulomb}^{(Rs)}(Q_{3})=G_{12}G_{23}G_{31}$. 
   
   To this purpose we use the NA44 data sample of three pion events 
   produced in S-Pb collisions at CERN \cite{na44}. 
The pions of this sample are identified and well separated charged particles.
All magnitudes of relative momenta $|{\bf k_{ij}}|$ are larger than 5 MeV due to experimental 
constraints and $Q_{3}$ is larger 
than 15 MeV.

We have calculated the Coulomb correction, i.e. $ K_{Coulomb}^{-1}(Q_{3})$, for radius values 
R=5 and 10 fm, i.e. the
range of interest for high energy heavy-ion physics, and compared the result
to the Riverside approximation, see Figure 1. For a detailed discussion of these data see the
NA44 publication \cite{na44}.

We find that already for a source
of radius 5 fm we need to take the detailed calculation into account as the difference
to the Riverside approximation is around 5-10\%, furthermore the inset clarifies that
this difference is $Q_3$-dependent, resulting in 
significant changes in any parameter extracted to characterize the correlation function.
For a larger source radius, 10 fm, the difference between the estimates is even more pronounced.
We have checked that in the limit $R\rightarrow 0$ we recover the Riverside approximation.

In Figure 2 we show the result of a corresponding calculation for the
two-pion case \cite{prattbiy}, illustrating that a finite size of the source 
reduces the Coulomb effect, as estimated by the Gamow penetration factor,
much in the same way as it reduces the Coulomb effect in the three-pion case
from the Riverside approximation.

\begin{figure}[htb]
\centering
\epsfig{file=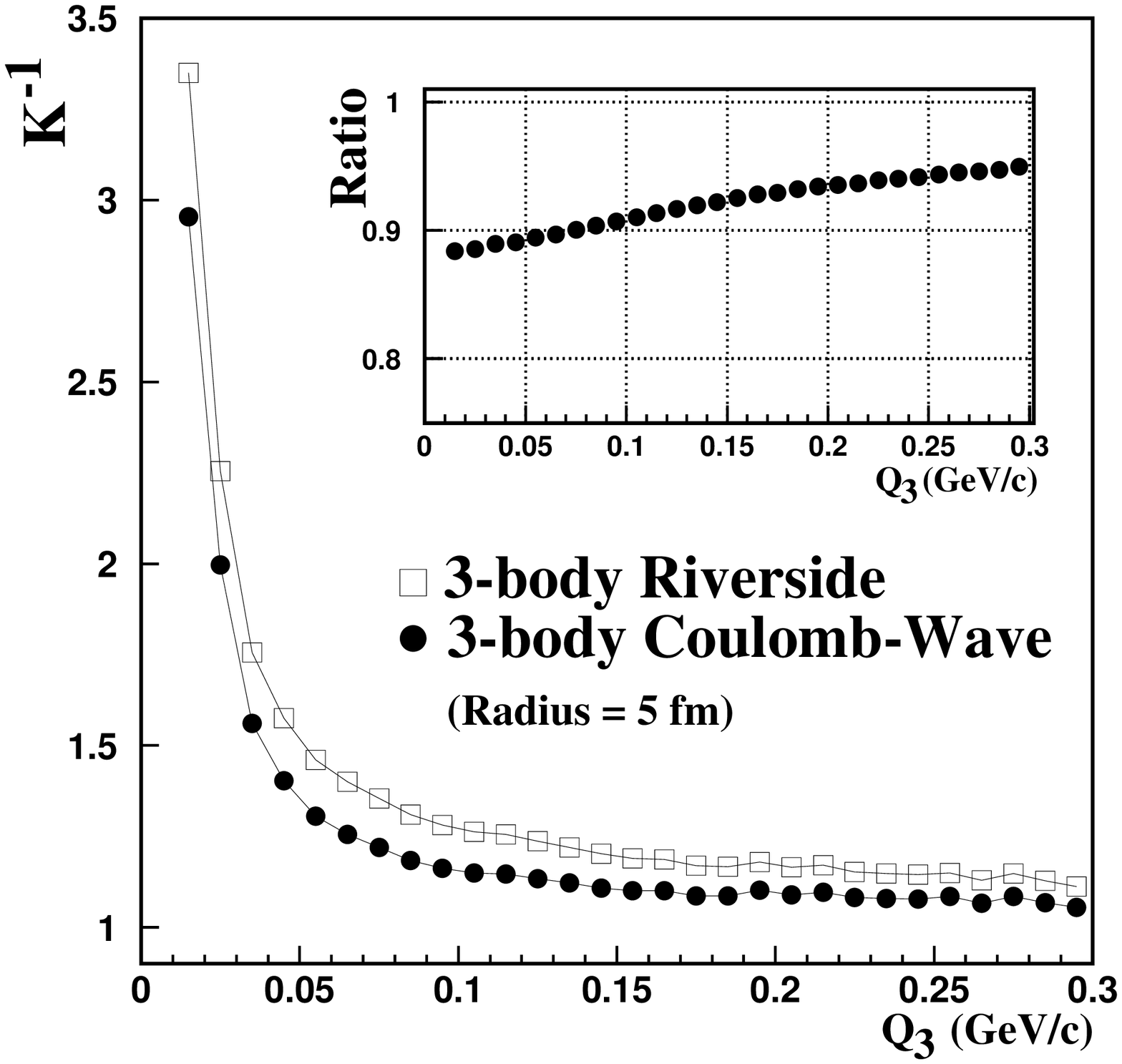,height=7.5cm}
\epsfig{file=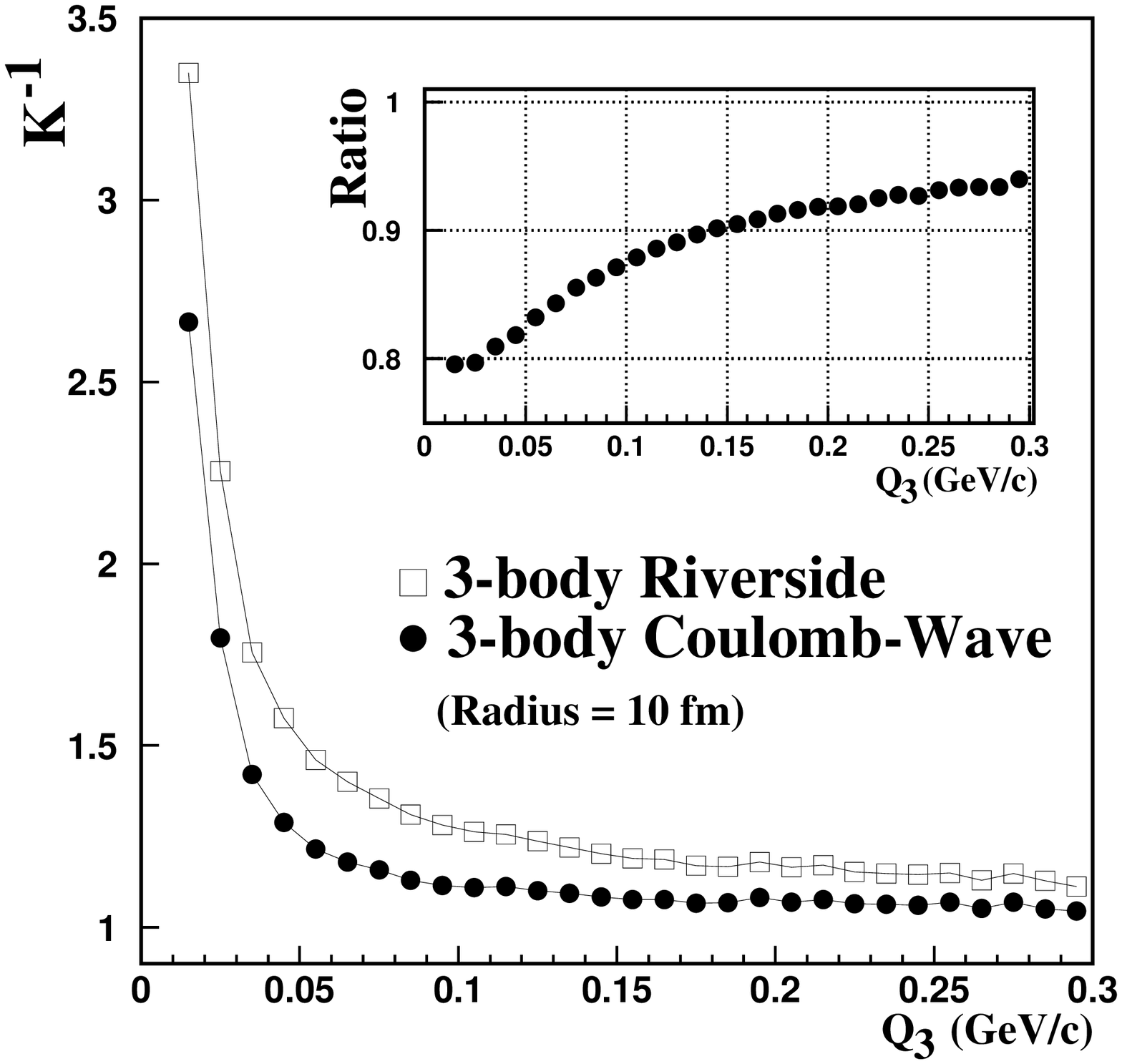,height=7.5cm}
\caption{The figures show the three-pion Coulomb correction factor $K_{Coulomb}^{-1}(Q_{3})$
as well as the Riverside approximation. In the upper panel, the input radius was 5 fm, while in
the lower, it was 10 fm. The insets display the ratio  Coulomb wavefunction integration
to the Riverside approximation. Lines are shown to guide the eye. }
\end{figure}

\begin{figure}[htb]
\centering
\epsfig{file=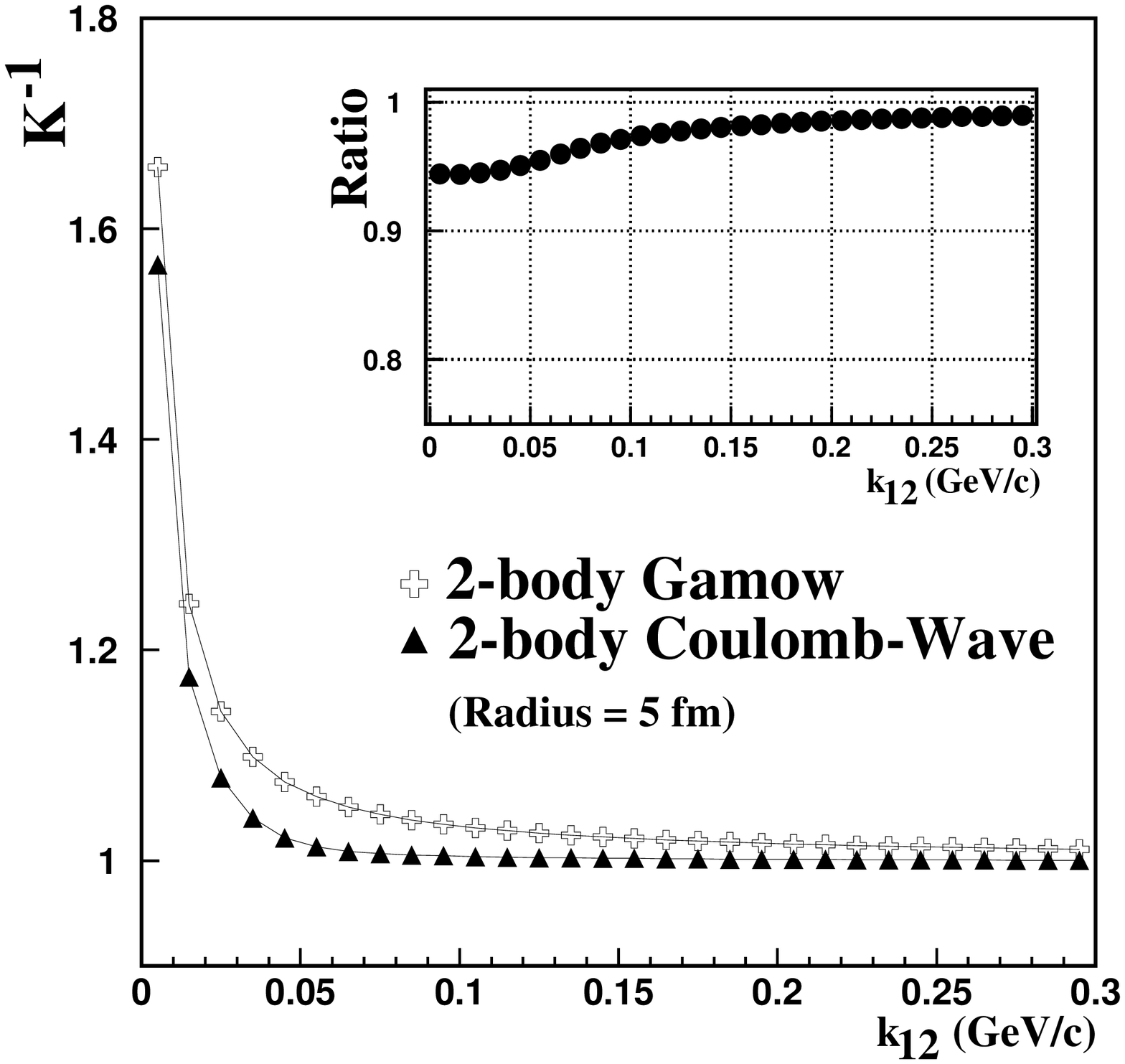,height=7.5cm}
\epsfig{file=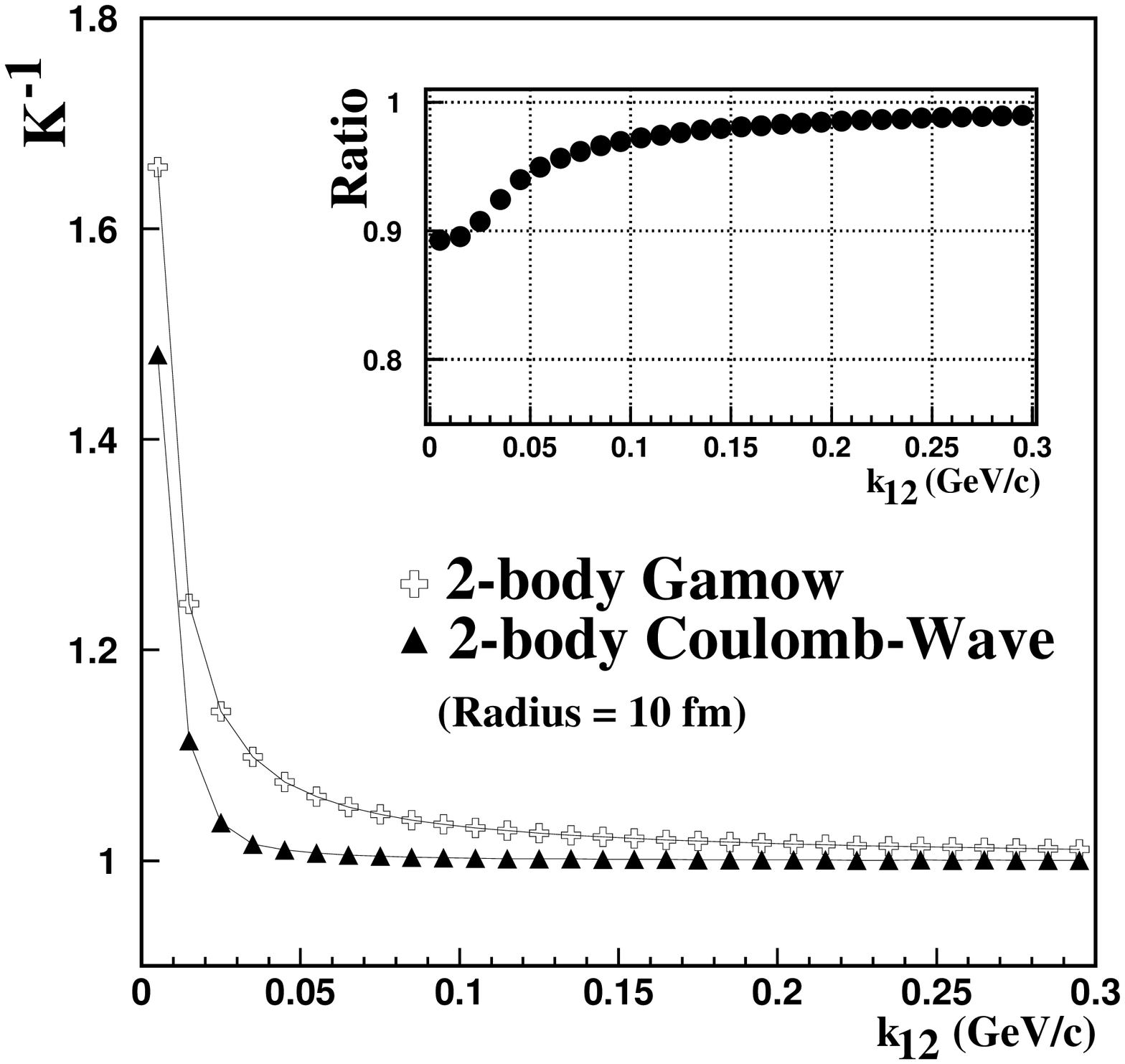,height=7.5cm}
\caption{The figures show the two-pion Coulomb correction factor $K_{Coulomb}^{-1}(k_{12})$
as well as the Gamow approximation. In the upper panel, the input radius was 5 fm, while in
the lower, it was 10 fm. The insets display the ratio  Coulomb wavefunction integration
to the Gamow approximation. Lines are shown to guide the eye. }
\end{figure}

\section{Discussion: Possible areas of future application}

Finally we highlight some of the physical problems in high-energy 
physics where the above results may have a future application
or where some measurements already indicate the possible influence of 
the many-body Coulomb distortions on the observables.

{\it 1)} In high-energy heavy ion physics nuclei
 are collided, that carry a large net positive charge and some of this 
charge may remain till the end of the reaction in the central collision
zone. Such a residual positive charge is expected to distort the
single-particle spectra of positively charged pions in different manner
than that of the single-particle spectra of negatively charged pions.
Namely, negatively charged pions are attracted to lower momenta,
while positively charged pions are repelled to higher values of the momenta.
This effect was seen in the NA44 data in Ref.\cite{na44sp} and
modeled in Ref. \cite{barz}.
This problem has a subtle three-body aspect, namely the
correlations between the positively and the negatively charged pions
induced by their Coulomb interaction which is also of long range, hence
in principle cannot be neglected. In this case, we have a charge
distribution $(Z, +, -)$ for which our wave function ansatz
(\ref{aswfn1}) can be utilized without the need of further
symmetrization to determine more precisely the magnitude of the residual
charge effects on the $N_1^{(-)} ({\bf k}) / N_1^{(+)}({\bf k})$
ratio of particle spectra. 

{\it 2)} If a residual net positive charge plays a r\^{o}le
on the distortion of charged particle spectra, its effects
must also be seen on the like-charged two-particle correlation functions.
For example, the two-particle intensity correlation functions
are symmetric in the absence of a central residual charge,
$C^{(++)} = C^{(--)}$, however, any net residual charge 
breaks the charge conjugation symmetry of the system,
$C^{(++)}_Z \ne C^{(--)}_Z$. This question could be properly studied
also in the framework of a Coulomb three-body problem with charges
$(Z,+,+)$ and $(Z,-,-)$ with the two-body symmetrization of the 
Coulomb relative wave-functions of the two identical bosons in this
system .

{\it 3)} The two-particle unlike-sign correlation function
is also distorted by the Coulomb-effects of the residual charge.
If we order the observed particles (pions), e.g., according to the 
magnitude of their momenta, and fix this order during the 
determination of $C_2^{(+-)}$ so that, e.g., the first particle is
always the one with the smaller momentum, we predict that
\begin{equation}
	C_2^{(+,-)}(\bk_1,\bk_2) \ne C_2^{(-,+)}(\bk_1,\bk_2)
	\qquad \mbox{if}\quad Z_{res} \ne 0
\end{equation}
This can be determined in a straigth-forward manner in experiments,
 either directly or by determining the charge-asymmetry of
the two-particle distribution functions,
\begin{equation}
	R^{(+-)} = \frac{N^{(+-)}(\bk_1,\bk_2) - N^{(-+)}(\bk_1,\bk_2)}
	{N^{(+-)}(\bk_1,\bk_2) + N^{(-+)}(\bk_1,\bk_2)}
	\label{rasym}
\end{equation}
Proper 3-body Coulomb calculations to estimate the magnitude of this
effect can be performed with the help of
the ansatz (\ref{aswfn1}).

{\it 4)}
Essentially, the method of {\it 3)} works also for the Coulomb correction due to
a net residual charge for unlike-particle correlation functions.
 The study of 
these correlation functions was proposed recently
by Lednicky and collaborators to access the temporal sequence of the
particle emission\cite{unlike}. This kind of physical information
is not directly accessible to like-particle correlation measurements,
and may provide a unique information on the dynamics of particle production.
Hence it is of great importance to determine experimentally the
value of the charge asymmetry parameter of the correlation
functions, Eq. (\ref{rasym}). If this parameter is non-vanishing,
three-body Coulomb-wave calculations have to be applied for unlike-particle 
correlations functions before determining
precisely the temporal sequence of particle emission from 
high-energy heavy ion experiments.

{\it 5)} In particle physics, Andersson and Ringn\'er recently suggested
to utilize the three-particle intensity correlation function
$C_3({\bf k}_1,{\bf k}_2,{\bf k}_3)$ as a test of the hadronization 
mechanism to see the longitudinal stretching of the string field
in the decay of quark--anti-quark jets\cite{bo}.
The method to estimate the Coulomb effects imbedded in this problem
 is outlined in the present paper.

 Although this list can be continued we stop here because the above
discussion is sufficient to show that the presented results are
rather useful for a broad range of physical problems in high-energy
particle and heavy ion physics.

\section{Summary}
On the basis of an explicit, analytically given form of the 
three-body Coulomb wave function that is 
correct in a large (asymptotic) region of three-body configuration space, 
we have developed a new method to systematically correct for 
explicit three-body Coulomb effects which is 
applicable to data analysis in a broad range of measurements 
in high-energy physics. The Riverside approximation has been 
established as a limiting case for 
vanishing source sizes. 

Specifically, we have worked out 
our approach for a system of three charged pions. 
There, we have numerically estimated that the 
Riverside approximation is not precise enough to 
determine the magnitude on the 5 -10 \% level
of the three-body Coulomb correction factor
in the correlation function, 
if the characteristic HBT radius parameter is 5 - 10 fm,
which is the range of interest in high-energy heavy ion physics.

\ack{ One of the authors (Cs. T.) would like to
thank Gy. Bencze for stimulating discussions. 
This research was partially supported by the  Hungarian National Science
Foundation under grant OTKA T026435.

The results presented are the product of a workshop on the Coulomb 
three-body problem in high energy physics held in Budapest September 21-25 1998. We are very grateful
for that intense and productive but at the same time relaxed week.
Support from the Swedish Research Council is acknowledged.}

\end{document}